\documentclass[12pt]{article}
\pdfoutput=1
\usepackage{subfigure}
\usepackage{amssymb,amsmath}
\usepackage{graphicx}
\usepackage{color}
\usepackage{cancel}
\usepackage[colorlinks=true
,urlcolor=blue
,citecolor=blue
,linkcolor=blue
,pagecolor=blue
,linktocpage=true
,pdfproducer=medialab
]{hyperref}
\usepackage[a4paper,width=15.2cm]{geometry}

\usepackage{ulem}
\usepackage[section]{placeins}

\makeatletter \renewcommand{\@dotsep}{10000} \makeatother
\def\be{\begin{equation}}
\def\ee{\end{equation}}
\def\bea{\begin{eqnarray}}
\def\eea{\end{eqnarray}}
\def\bi{\begin{itemize}}
\def\ei{\end{itemize}}

\usepackage[nodisplayskipstretch]{setspace}

\newcommand{\beq}{\begin{equation}}
\newcommand{\eeq}{\end{equation}}

\newcommand{\eval}{\biggr\rvert}
\begin{document}
%Remove date before submitting to arXi

\date{\today}

\begin{center}
{\Large\bf 
Least Fine-Tuned $\mathbf{U(1)}$ Extended SSM } 
\end{center}

\begin{center}

{\Large
Ya\c{s}ar Hi\c{c}y\i lmaz$^{a}$, Levent Solmaz$^{a}$, \\
\vspace{0.3cm}
\c{S}\"{u}kr\"{u} Hanif Tany\i ld\i z\i $^{a}$~ and Cem Salih \"{U}n$^{b}$
}

\vspace{0.75cm}

{\it \hspace{-2.4cm}$^a$
Department of Physics, Bal\i kesir University, 10145, Bal\i kesir Turkey} \\

\vspace{0.2cm}

 {\it \hspace{-3.1cm}$^b$
Department of Physics, Uluda{\~g} University, 16059, Bursa, Turkey
}
%

%\vspace{1.5cm}
\section*{Abstract}
\end{center}
\noindent
We consider the Higgs boson mass in a class of {the UMSSM} models in which the MSSM gauge group is extended by an additional $U(1)'$ group. {Implementing the universal boundary condition at the GUT scale we target phenomenologically interesting regions of UMSSM where the necessary radiative contributions to the lightest CP-even Higgs boson mass are significantly small and LSP is always the lightest neutralino}. We find that the smallest amount of radiative contributions to the Higgs boson mass is about 50 GeV in UMSSM, this result is much lower than that obtained in the MSSM framework, which is around 90 GeV. Additionally, we examine the Higgs boson properties in these models in order to check whether if it can behave similar to the SM Higgs boson under the current experimental constraints. We find that enforcement of smaller radiative contribution mostly restricts the $U(1)'$ breaking scale as $v_{S} \lesssim 10$ TeV. Besides, such low contributions demand $h_S \sim 0.2 - 0.45$. Because of the model dependency in realizing these radiative contributions  $\theta_{E_{6}} < 0$ are more favored, if one seeks for the solutions consistent with the current dark matter constraints.  As to the mass spectrum, we find that stop and stau can be degenerate with the LSP neutralino in the range from 300 GeV to 700 GeV; however, the dark matter constraints restrict this scale as $m_{\tilde{t}}, m_{\tilde{\tau}} \gtrsim 500$ GeV. Such degenerate solutions also predict stop-neutralino and stau-neutralino coannihilation channels, which are effective to reduce the relic abundance of neutralino down to the ranges consistent with the current dark matter observations. {Finally, we discuss the effects of heavy $M_{Z'}$ in the fine-tuning. Even though the radiative contributions are significantly low, the required fine-tuning can still be large. We comment about reinterpretation of the fine-tuning measure in the UMSSM framework, which can yield efficiently low results for the fine-tuning the electroweak scale.}

\newpage

%%%%%%%%%%%%%%%%%%%%%%%%%%%%%%%%%%%%%%%%%%%%%%%%%%%%%%%%%%%%
\renewcommand{\thefootnote}{\arabic{footnote}}
\setcounter{footnote}{0}

%%%%%%%%%%%%%%%%%%%%%%%%%%%%%%%%%%%%%%%%%%%%%%%%%%%%%%%%%%%%%

%\baselineskip 36pt
% Main body
%%%%%%%%%%%%%%%%%%%%%%%%%%
%\baselineskip 18pt
%%%%%%%%%%%%%%%%%%%%%%%%%%

%%%%%%%%%%%%%%%%%%%%%%%%%%%%
\section{Introduction}
\label{sec:intro}

Even though the minimal supersymmetric extension of the Standard Model (MSSM) is compatible with the Higgs boson of mass about 125 GeV as observed by the ATLAS \cite{Aad:2012tfa} and CMS \cite{Chatrchyan:2012xdj} collaborations, it brings back the naturalness and fine-tuning discussions \cite{Baer:2012mv}, since it requires very heavy stop quarks or large trilinear scalar interaction couplings \cite{Heinemeyer:2011aa}. Besides, null results from the experimental analyses for direct signals of supersymmetric particles also have lifted up the mass bounds on the supersymmetric particles.

The experiments conducted at the Large Hadron Collider (LHC) mostly bound the colored supersymmetric particles such as stop and gluino. Although these particles have nothing to do with the fine-tuning assertions at tree-level, they are linked to the electroweak (EW) sector when the universal boundary conditions are applied at the grand unification scale ($M_{{\rm GUT}}$). In this case, the mass bound on gluino can be set as $m_{\tilde{g}} \geq 1.8$ TeV \cite{TheATLAScollaboration:2015cyl} also leads to heavy Bino and Wino when $M_{1}=M_{2}=M_{3}=M_{1/2}$ at $M_{{\rm GUT}}$, which yield large fine-tuning at the EW scale. The mass bound on the stop differs depending on the decay channels of stops, and it can be as low as about 230 GeV, when it decays into a neutralino and a charm quark \cite{TheATLAScollaboration:2013aia}. However, in the case of such light stop solutions, the Higgs boson mass requirement yields large trilinear scalar interaction coupling ($A_{t}$).  Although an acceptable amount of fine-tuning can be realized even if the stop is heavy, recent studies \cite{Demir:2014jqa} show that the mixing in the stop sector, which is proportional to $A_{t}$, raises the fine-tuning measurements, since $A_{t}$ significantly enhances the soft supersymmetry breaking (SSB) mass of $H_{u}$ ($m_{H_{u}}$) at loop-level. 

The large fine-tuning results obtained within the MSSM framework are based on the fact that MSSM yields inconsistently low mass for the Higgs boson at tree-level, and one needs to utilize the loop corrections to obtain large radiative contributions to the Higgs boson. Since the particles in the first two families negligibly couple to the Higgs boson, such corrections can only come from the third family supersymmetric particles. On the other hand, couplings of the Higgs boson with sbottom and stau can easily destabilize the Higgs potential, and hence, the Higgs potential stability condition allows only minor contributions to the Higgs boson mass from these particles \cite{Carena:2012mw}. After all, MSSM has only the stop sector to provide large enough radiative contributions to the Higgs boson mass, which needs to have both heavy stops and large mixing in the stop sector.  In this context, the supersymetric models with extra sectors, which couple to the MSSM Higgs doublets (especially to $H_{u}$) can relax the pressure on the stop sector, and alleviate the large fine-tuning issue, even if one applies universal boundary conditions at $M_{{\rm GUT}}$ (see, for instance \cite{Hicyilmaz:2016kty}). 

In this paper, we consider the models, which extend the MSSM group with an extra $U(1)$ symmetry, hereafter UMSSM for short. In this extension of MSSM, all particles, including $H_{u}$ and $H_{d}$, have non-trivial charges under the extra $U(1)$ gauge group, and hence, the Higgs mass receives extra contributions from the new sector at even tree-level, which results in reducing the necessary amount of the radiative corrections to the Higgs boson mass. It is interesting to probe the necessary amount of loop corrections and fine-tuning issues within such gauge extended supersymmetric models whether it can be smaller than MSSM or not. The rest of the paper is organized as follows. Section \ref{sec:model} briefly discusses the general properties and the particle content of the UMSSM. The Higgs boson mass is discussed in Section \ref{sec:hmass}. After we summarize our scanning procedure and the experimental constraints employed in our analyses in Section \ref{sec:scan}, we first consider the profile of the Higgs boson compared to the SM Higgs boson and related decay channels of it, in Section \ref{sec:hprofile}. After highlighting the solutions which can yield Higgs boson with similar properties to that in SM, we discuss how much low radiative corrections can be acceptable under the current Higgs boson observations in Section \ref{sec:R}. We discuss about fine-tuning in connection with  low amounts of the radiative contributions in Section \ref{sec:FT}, and finally, we summarize and conclude our findings in Section \ref{sec:conc}.

\section{Model Description and Particle Content}
\label{sec:model}

A general extension of MSSM by a $U(1)$ group can be realized from an underlying grand unified theory (GUT) involving a gauge group larger than $SU(5)$ (for a detailed description of the model, see \cite{Barr:1985qs,Langacker:1998tc}). In this context, one can have a significant freedom in choice of the extra $U(1)$ group, when it is obtained through the breaking pattern of the exceptional group $E_{6}$ given as 

\begin{equation}
E_{6}\rightarrow SO(10)\times U(1)_{\psi}\rightarrow SU(5)\times U(1)_{\chi}\times U(1)_{\psi}\rightarrow G_{{\rm MSSM}}\times U(1)' 
\label{E6breaking}
\end{equation}
where $G_{{\rm MSSM}}=SU(3)_{c}\times SU(2)_{L}\times U(1)_{Y}$ is the MSSM gauge group, and $U(1)'$ can be expressed as a general mixing of $U(1)_{\psi}$ and $U(1)_{\chi}$ as  

\begin{equation}
U(1)'=\cos \theta_{E_{6}}U(1)_{\chi}+\sin\theta_{E_{6}}U(1)_{\psi}.
\label{Umixing}
\end{equation}

If the matter particles reside in $\mathbf{27}-$dimensional representation of $E_{6}$, its decomposition yields additional vector-like families denoted by $\Delta$ and $\bar{\Delta}$ \cite{Langacker:1998tc}. These additional vector-like families are crucial in anomaly cancellation in UMSSM. The presence of these vector-like fields does not break the gauge coupling unification at $M_{{\rm GUT}}$, while they change the $\beta -$functions of the MSSM gauge couplings to $(b_{1},b_{2},b_{3}=\frac{48}{5},4,0)$ \cite{Falck:1985aa}. Including these vector-like fields the superpotential can be written in UMSSM as follows:

\begin{equation}
W = Y_{u}\hat{Q}\hat{H}_{u}\hat{U}^{c}+Y_{d}\hat{Q}\hat{H}_{d}\hat{D}^{c}+Y_{e}\hat{L}\hat{H}_{d}\hat{E}^{c}+h_{S}\hat{S}\hat{H}_{d}\hat{H}_{u}+h_{\Delta}\hat{S}\hat{\Delta}\hat{\bar{\Delta}}.
\label{suppot1}
\end{equation}

where $\hat{Q}$ and $\hat{L}$ denote the left-handed chiral superfields for the quarks and leptons, while $\hat{U}^{c}$, $\hat{D}^{c}$ and $\hat{E}^{c}$ stand for the right-handed chiral superfields of u-type quarks, d-type quarks and leptons, respectively. $H_{u}$ and $H_{d}$ MSSM Higgs doublets and $Y_{u,d,e}$ are their Yukawa couplings to the matter fields. In addition to the MSSM content and the vector-like fields $\Delta$ and $\bar{\Delta}$, $\hat{S}$ also denotes a chiral superfield. This field is preferably a singlet under the MSSM group and its vacuum expectation value (VEV) is responsible for the breaking of $U(1)^\prime$ symmetry. The MSSM particles are also non-trivially charged under $U(1)_{\chi}$ and $U(1)_{\psi}$, and the invariance under $U(1)^\prime$ requires an appropriate charge assignment for the MSSM fields. Table \ref{charges} displays the charge configurations for $U(1)_{\psi}$ and $U(1)_{\chi}$ models. When these two gauge groups mix each other as given in Eq.(\ref{Umixing}), the following equation describes the resultant charge of the MSSM particles:

\begin{equation}
Q^{i}=Q^{i}_{\chi}\cos\theta_{E_{6}}+Q^{i}_{\psi}\sin\theta_{E_{6}}.
\label{Chmixing}
\end{equation}

\begin{table}[ht!]
\setstretch{1.5}
\centering
\begin{tabular}{|c|c|c|c|c|c|c|c|c|c|c|}
\hline
 Model & $\hat{Q}$ & $\hat{U}^{c}$ & $\hat{D}^{c}$ & $\hat{L}$ & $\hat{E}^{c}$ & $\hat{H}_{d}$ & $\hat{H}_{u}$ & $\hat{S}$ & $\Delta$ & $\bar{\Delta}$\\
 \hline
$ 2\sqrt{6}~U(1)_{\psi}$ & 1 & 1 & 1 &1 &1 & -2 & -2 & 4 & -2 & -2\\
\hline 
$ 2\sqrt{10}~U(1)_{\chi}$ & -1 & -1 & 3 & 3 & -1 & -2 & 2 & 0 &2 & -2\\
\hline  
\end{tabular}
\caption{Charge assignments for the fields in several models.}
\label{charges}
\end{table}
Note that the bilinear mixing of the MSSM Higgs doublets, given as $\mu H_{d}H_{u}$, is forbidden in the superpotential given in Eq.(\ref{suppot1}) by the invariance under $U(1)'$, and it is induced effectively by the VEV of $\hat{S}$ as $\mu = h_{S}v_{S}/\sqrt{2}$, where $v_{S}\equiv \langle S \rangle$. Besides, the VEV of $S$ along with its Yukawa coupling $h_{\Delta}$ is responsible for the masses of $\Delta$ and $\bar{\Delta}$. If $h_{\Delta}$ is set to large values, then these vector-like fields happen to be so heavy that they decouple at a high energy scale \cite{Hebbar:2016gab}. Since they interact only with $S$, they contribute to the mass spectrum through higher loop levels, which are strongly suppressed by their heavy masses.

Extending the gauge group of MSSM also enlarges the particle content with new particles, which interfere the low scale phenomenology. In addition to the MSSM gauge fields, there also exists a new gauge boson ($Z'$) and its supersymmetric partner ($\tilde{B}'$) associated with $U(1)'$ symmetry. The negative LEP results strictly constrain the $Z'$ mass  from below as $M_{Z'}/g' \geq 6$ TeV \cite{Cacciapaglia:2006pk}, where $g'$ is the gauge coupling of $U(1)'$. Even though detailed analyses \cite{Accomando:2013sfa} can lower this bound, we consider only the solutions yielding heavy $Z'$.  The effect of heavy $Z'$ can be seen from its mass equation given as 

\begin{equation}
M_{Z'}^{2}=g'^{2}(Q^{2}_{H_{u}}v_{u}^{2}+Q^{2}_{H_{d}}v_{d}^{2}+Q^{2}_{S}v_{S}^{2})
\label{Zpmass}
\end{equation}
where $Q_{H_{u},H_{d},S}$ denote the charges of these fields under $U(1)'$, and $v_{u,d,S}$ are their VEVs. Since the charges are fixed by the $U(1)'$ gauge group and $v_{u,d}$ are strictly constrained by the electroweak data as $\sqrt{v_{u}^{2}+v_{d}^{2}} \approx 246$ GeV, heavy $M_{Z'}$ leads to large $g'v_{S}$. Requiring the gauge coupling unification at $M_{{\rm GUT}}$ including $g'$, $v_{S}$ needs to be large to provide heavy $Z'$; hence, the breaking of $U(1)'$ symmetry cannot happen at energy scales below a few TeV. In addition, $Z'$ can also mix with the electroweak neutral gauge boson $Z$, and the diagonalization of their mass matrix yields the following mass eigenstates for these gauge bosons

\begin{equation}
M^{2}_{Z,Z'}=\frac{1}{2}\left[ M_{Z}^{2}+M_{Z'}^{2}\mp\sqrt{(M_{Z}^{2}-M_{Z'}^{2})^{2}+4\delta_{Z-Z'}}   \right]
\label{ZZprime}
\end{equation}
where $\delta_{Z-Z'}$ refers to the mixing between $Z$ and $Z'$. Even though $Z'$ can, in principle, interfere in the electroweak processes through Eq.(\ref{ZZprime}), $M_{Z'}\sim \mathcal{O}(TeV)$ strongly suppresses such mixing; therefore, $Z-$boson is realized more or less identical to the MSSM electroweak neutral gauge boson. Despite the heavy mass bound on $Z'$, there is no specific bound on the mass of its supersymmetric partner $\tilde{B}'$, and it is possible to realize $\tilde{B}'$ mass as low as about 100 GeV \cite{Khalil:2015wua}.

Another extra particle introduced is $S$, which is responsible for the $U(1)'$ symmetry breaking. If its coupling ($h_{\Delta}$) to the vector-like fields $\Delta$ and $\bar{\Delta}$ is set to be large, this coupling can drive $m_{S}$ down through the renormalization group (RG) evolution, and hence $S$ can be realized with a TeV scale mass at the low scale. The largest impact of the $U(1)'$ symmetry is realized in the neutralino sector. The electroweak symmetry breaking in MSSM mixes the neutral gauginos and Higgsinos to each other. Similarly, the breaking of $U(1)'$ symmetry allows $\tilde{B}'$ and the fermionic partner of $S$ to mix with the MSSM neutral gauginos and higgsinos; hence, they take place in forming the neutralino mass eigenstates. In this context, UMSSM yields six neutralinos at the low scale, and if $\tilde{B}'$ can be light, it might significantly change nature of the neutralino LSP, if it is considered as a dark matter (DM) candidate. 

Since $U(1)'$ symmetry does not introduce any charged particle, the chargino sector remains intact, and hence UMSSM and MSSM bear the same chargino structures. However, since the $\mu-$parameter is induced effectively, UMSSM may yield different Higgsino mass scale from that realized in the MSSM framework, which can change nature of the lightest chargino.

In addition to the superpotential, the SSB Lagrangian is given as

\begin{equation*}
-\mathcal{L}_{\cancel{SUSY}}=m_{\tilde{Q}}^2|\tilde{Q}|^2+m_{\tilde{U}}^2|\tilde{U}|^2+m_{\tilde{D}}^2|\tilde{D}|^2+m_{\tilde{E}}^2|\tilde{E}|^2+m_{\tilde{L}}^2|\tilde{L}|^2
\end{equation*}

\begin{equation*}
+m_{H_{u}}^2|H_u|^2+m_{H_{d}}^2|H_d|^2+m_S^2|S|^2 + m_{\tilde{\Delta}}^{2}|\Delta|^{2}+\sum_a M_a\lambda_a\lambda_a
\end{equation*}

\begin{equation}
+\left( A_SY_SSH_u\cdot H_d+A_tY_t\tilde{U}^c\tilde{Q}\cdot H_u+A_bY_b\tilde{D}^c\tilde{Q}\cdot H_d+A_{\tau}Y_b\tilde{L}^c\tilde{e}\cdot H_d+h.c. \right)
\end{equation}
where $m_{\tilde{Q}}$, $m_{\tilde{U}}$, $m_{\tilde{D}}$, $m_{\tilde{E}}$, $m_{\tilde{L}}$,$m_{H_{u}}$, $m_{H_{d}}$, $m_{\tilde{S}}$ and $m_{\tilde{\Delta}}$ are the mass matirces of the particles identified with the subindices, while $M_{a}\equiv M_{1},M_{2},M_{3},M_{4}$ stand for the gaugino masses. $A_{S}$, $A_{t}$, $A_{b}$ and $A_{\tau}$ are the trilinear scalar interction couplings, and they are factorized in terms of the Yukawa couplings; and hence, we consider only the third family MSSM particles, since the first two families have negligible Yukawa couplings with the Higgs doublets. Even though the number of free parameters seems too many, the emergence of $SO(10)$ and/or $SU(5)$ allows to implement a set of boundary conditions among these parameters at $M_{{\rm GUT}}$. In this paper, we implemented the following universal boundary conditions

\begin{equation}\hspace{-1.5cm}
\setstretch{1.5}
\begin{array}{ll}
m_{0} & = m_{\tilde{Q}}=m_{\tilde{U}}=m_{\tilde{D}}=m_{\tilde{E}} = m_{\tilde{L}}=m_{\tilde{Q}}=m_{H_{u}}=m_{H_{d}}=m_{\tilde{S}}=m_{\tilde{\Delta}}  \\

M_{1/2} &=  M_{1}=M_{2}=M_{3}=M_{4}  \\

A_{0} & =  A_{t}=A_{b}=A_{\tau}=A_{S}=A_{\Delta}. 
\end{array}
\end{equation}

\section{Higgs Boson Mass in UMSSM}
\label{sec:hmass}

As mentioned before, MSSM predicts inconsistently light Higgs boson mass at tree-level, and hence it needs large radiative corrections in order to satisfy the Higgs boson mass constraint. On the other hand, UMSSM provides new contributions to the Higgs boson mass at tree-level, and hence the radiative corrections may not need to be very large. In our model, the tree-level Higgs boson mass can be obtained by the tree-level Higgs potential expressed as 

\begin{equation}
V^{{\rm tree}}=V_{F}^{{\rm tree}}+V_{D}^{{\rm tree}}+V_{\cancel{SUSY}}^{{\rm tree}}
\end{equation}
with 

\begin{equation}
\setstretch{2.0}
\begin{array}{ll}
V_{F}^{{\rm tree}} & = | Y_{S} |^{2} \left[| H_{u}H_{d}|^{2} + | S |^{2}\left( | H_{u}|^{2}+| H_{d}|^{2}  \right)   \right] \\
V_{D}^{{\rm tree}} & = \dfrac{g_{1}^{2}}{8}\left( | H_{u}|^{2}+| H_{d}|^{2}  \right)^{2}+\dfrac{g_{2}^{2}}{2}\left( |H_{u}|^{2}|H_{d}|^{2}-|H_{u}H_{d}|^{2}  \right) \\

& + \dfrac{g'^{2}}{2}\left( Q_{H_{u}}|H_{u}|^{2}+Q_{H_{d}}|H_{d}|^{2}+Q_{S}|S|^{2}  \right) \\

V_{\cancel{SUSY}}^{{\rm tree}} & = m^{2}_{H_{u}}|H_{u}|^{2}+m_{H_{d}}^{2}|H_{d}|^{2}+m_{S}^{2}|S|^{2}+\left(A_{S}Y_{S}SH_{u}H_{d}+h.c. \right),
\end{array}
\end{equation}
which yields the following tree-level mass for the lightest CP-even Higgs boson mass:

\begin{equation}
m_h^2=M_Z^2\cos^22\beta+\left(v_u^2+v_d^2\right)\left[\frac{h_S^2\sin^22\beta}{2}+g_{Y^\prime}^2\left(Q_{H_{u}}\cos^2\beta+Q_{H_{d}}\sin^2\beta\right)\right].
\label{h0mass}
\end{equation}

\begin{figure}[h!]
\subfigure{\includegraphics[scale=0.4]{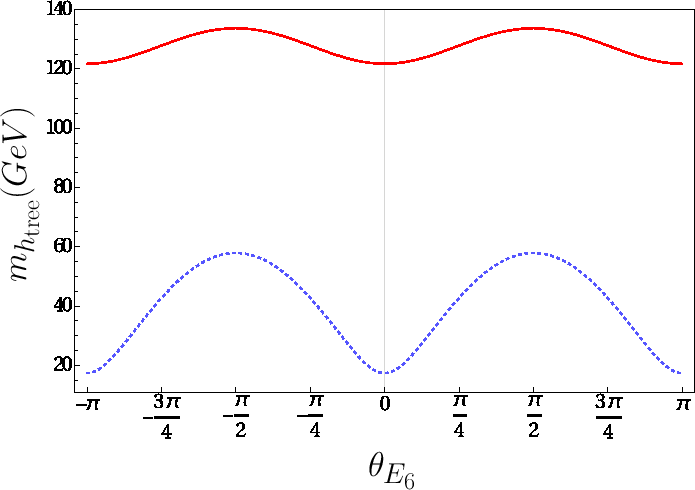}}
\subfigure{\includegraphics[scale=0.4]{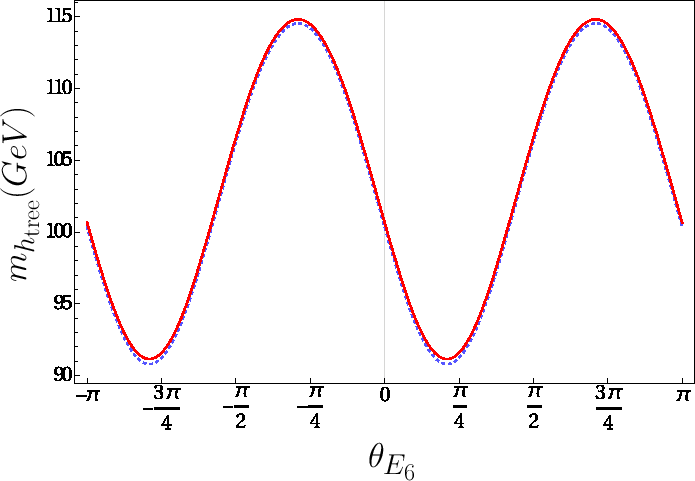}}
\caption{Model dependency of the tree-level Higgs mass in correlation with $\theta_{E_{6}}$ for $\tan\beta = 1$ (left) and $\tan\beta =30$ (right). The dotted blue curves in both panels represent the tree-level Higgs boson mass when $h_{S}=0.1$, while the solid red curves are obtained for $h_{S}=0.7$. }
\label{fig:mhtree}
\end{figure}

The first term in Eq.(\ref{h0mass}) is the MSSM prediction for the lightest CP-even Higgs boson mass, and it can barely reach to about 90 GeV; therefore, one needs at least to have radiative corrections of about 90 GeV in the best case. On the other hand, the second term in Eq.(\ref{h0mass}) provided by UMSSM can alleviate the need for the large radiative corrections to the Higgs boson mass. Apart from the couplings $h_{S}$ and $g_{Y'}$, the tree-level Higgs boson mass also depends on the charges of $H_{u}$ and $H_{d}$ under the $U(1)'$. These charges exhibit model dependency, since they vary as functions of the mixing angle between different $U(1)$ groups as expressed in Eq.(\ref{Umixing}). Hence, the upper bound for the tree-level Higgs boson mass can change from one model to another as it can be seen from Figure \ref{fig:mhtree}, where the model dependency of the tree-level Higgs boson mass is represented in correlation with $\theta_{E_{6}}$ for  $\tan\beta = 1$ (left) and $\tan\beta =30$ (right). The dotted blue curves in both panels represent the tree-level Higgs boson mass when $h_{S}=0.1$, while the solid red curves are obtained for $h_{S}=0.7$. The dotted blue curve in the left panel shows that the Higgs boson can only be as heavy as about 60 GeV at tree-level, when $h_{S}=0.1$ and $\tan\beta = 1$. On the other hand, the upper bound obtained for the tree-level Higgs boson mass in UMSSM can drastically raise up to $\sim 140$ GeV, when $h_{S}=0.7$, as shown with the red curve in the left panel. The sensitivity to $h_{S}$ almost disappears when $\tan\beta =30$. The right panel shows that the largest tree-level Higgs boson mass can be realized as about 115 GeV. Since the dotted blue curve and the solid red curve overlap each other, the effect from $h_{S}$ on the Higgs boson mass is quite tiny and negligible, even though it is varied from 0.1 to 0.7. It is because $\sin 2\beta \sim 0$ when $\tan\beta$ is large, which suppresses the contribution from $h_{S}$ to the Higgs boson mass. These values are predicted when UMSSM is constrained at the GUT scale, which yield $h_S \simeq 0.7$ at most. As is known, if UMSSM is considered at the low energy scale; then, the tree-level Higgs boson mass can be obtained as heavy as about 180 GeV \cite{Sert:2010ma}.

Although they do not take part in tree-level Higgs boson mass prediction, the SUSY particles contribute to the Higgs boson mass through loops. Even if a solution can yield heavy Higgs boson at tree-level, the SUSY particle spectrum for such a solution can still provide large radiative corrections to the Higgs boson mass, so the solution can be excluded since it predicts inconsistently heavy Higgs boson mass. The radiative corrections to the Higgs boson mass can be obtained by using the effective potential method in which the effective Higgs potential can be expressed as 

\begin{equation}
V^{{\rm eff}}=V^{{\rm tree}}+\Delta V~, \hspace{0.3cm}{\rm with}\hspace{0.3cm} \Delta V = \frac{1}{64\pi^{2}}STr \left[\mathcal{M}^{4}\left(\log \frac{\mathcal{M}^{2}}{\Lambda^{2}} -\frac{3}{2}\right) \right]
\end{equation}
where $STr=\sum_{J}(-1)^{2J}(2J+1)Tr$ stands for the supertrace, and it gives a factor of -12 for quarks and 6 for squarks (for a detailed discussion about the effective potential see \cite{Carena:1995wu}). Following the effective potential approach the Higgs boson mass with one-loop corrections can be obtained as \cite{Demir:1998dk}

\begin{equation}
\setstretch{2.0}
\begin{array}{l}
m_{h_{\textsl{loop}}}^2=m_h^2+\beta_{y_t}\left[\left(\mu\cos\beta+A_t\sin\beta\right)^2+4S_{t\tilde{t}}m_t^2\right] \\
\Delta m_{h}^{2}\equiv m_{h_{{\rm loop}}}^{2}-m_{h_{{\rm tree}}}^{2}=\beta_{y_t}\left[\left(\mu\cos\beta+A_t\sin\beta\right)^2+4S_{t\tilde{t}}m_t^2\right]
\end{array}
\label{h0loop}
\end{equation}
where $m_{h}$ is the tree-level mass of the Higgs boson as given in Eq.(\ref{h0mass}), $\beta_{y_{t}}=(3/16\pi^{2})y_{t}^{2}$, and $S_{t\tilde{t}} = \log(m_{\tilde{t}_{1}}m_{\tilde{t}_{2}}/m_{t}^{2})$ encodes the loop effects of the top-stop mass splitting. Even though there are some other sources for the radiative contributions to the Higgs boson mass from sbottom, stau, neutralino etc., such contributions are rather minor, and as in the case of MSSM, also in UMSSM the radiative corrections to the Higgs boson mass rely mainly on the stop sector.

\section{Scanning Procedure and Constraints}
\label{sec:scan}

We have employed SPheno 3.3.3 package \cite{Porod:2003um} obtained with SARAH 4.5.8 \cite{Staub:2008uz}. In this package, the weak scale values of the gauge and Yukawa couplings present in UMSSM are evolved to the unification scale $M_{{\rm GUT}}$ via the renormalization group equations (RGEs). $M_{{\rm GUT}}$ is determined by the requirement of the gauge coupling unification through their RGE evolutions. Note that we do not strictly enforce the unification condition $g_{1}=g_{2}=g_{3}=g'_{Y}$ at $M_{{\rm GUT}}$ since a few percent deviation from the unification can be assigned to unknown GUT-scale threshold corrections \cite{Hisano:1992jj}. Such corrections are rather effective in $g_{3}$, and hence the unification condition can be relaxed up to $3\%$ deviation in $g_{3}$. With the boundary conditions given at $M_{{\rm GUT}}$, all the SSB parameters along with the gauge and Yukawa couplings are evolved back to the weak scale. During our numerical investigation, we have performed random scans over the following parameter space

\begin{equation}
\setstretch{1.2}
\begin{array}{ccc}
0 \leq & m_{0} & \leq 5 ~{\rm (TeV)} \\
0 \leq & M_{1/2} & \leq 5 ~{\rm (TeV)} \\
1.2 \leq & \tan\beta & \leq 50 \\
-3 \leq & A_{0}/m_{0} & \leq 3 \\
-10 \leq & A_{S} & \leq 10 ~{\rm (TeV)} \\
1 \leq & v_{S} & \leq 25~{\rm (TeV)} \\
0 \leq & h_{S} & \leq 0.7 \\
-\dfrac{\pi}{2} \leq & \theta_{E_{6}} & \leq \dfrac{\pi}{2}
\end{array}
\label{paramSP}
\end{equation}  
where $m_{0}$ is the universal SSB mass term for all the scalar fields including $H_{u}$, $H_{d}$, $S$ fields, and similarly $M_{1/2}$ is the universal SSB mass term for the gaugino fields including one associated with $U(1)'$ gauge group. $\tan\beta=\langle v_{u} \rangle / \langle v_{d} \rangle$ is the ratio of VEVs of the MSSM Higgs doublets, $A_{0}$ is the SSB trilinear scalar interaction term. Similarly, $A_{h_{S}}$ is the SSB interaction between the $S$ and $H_{u,d}$ fields, which is varied free from $A_{0}$ in our scans. Finally, $v_{S}$ denotes the VEV of $S$ fields which indicates the $U(1)'$ breaking scale. Recall that the $\mu-$term of MSSM is dynamically generated such that $\mu = h_{S}v_{S}/\sqrt{2}$. Its sign is assigned as a free parameter in MSSM, since REWSB condition can determine its value but not sign. On the other hand, in UMSSM, it is forced to be positive by $h_{S}$ and $v_{S}$. Finally, we set the top quark mass to its central value ($m_{t} = 173.3$ GeV) \cite{Group:2009ad}. Note that the sparticle spectrum is not too sensitive in one or two sigma variation in the top quark mass \cite{Gogoladze:2011db}, but it can shift the Higgs boson mass by $1-2$ GeV  \cite{Gogoladze:2011aa}.

The requirement of radiative electroweak symmetry breaking (REWSB) \cite{Ibanez:1982fr} puts an important theoretical constraint on the parameter space. Another important constraint comes from the relic abundance of the stable charged particles \cite{Nakamura:2010zzi}, which excludes the regions where charged SUSY particles such as stau and stop become the lightest supersymmetric particle (LSP). In our scans, we allow only the solutions for which one of the neutralinos is the LSP and REWSB condition is satisfied.

In scanning the parameter space, we use our interface, which employs Metropolis-Hasting algorithm described in \cite{Belanger:2009ti}. After collecting the data, we impose the mass bounds on all the sparticles \cite{Agashe:2014kda}, and the constraint from the rare B-decays such as $B_{s}\rightarrow \mu^{+}\mu^{-}$ \cite{Aaij:2012nna}, $B_{s}\rightarrow X_{s}\gamma$ \cite{Amhis:2012bh}, and $B_{u}\rightarrow \tau \nu_{\tau}$ \cite{Asner:2010qj}. In addition,  the WMAP bound \cite{Hinshaw:2012aka} on the relic abundance of neutralino LSP within $5\sigma$ uncertainty. Note that the current results from the Planck satellite \cite{Ade:2015xua} allow more or less a similar range for the DM relic abundance within $5\sigma$ uncertainty, when one takes the uncertainties in calculation. These experimental constraints can be summarized as follows:

\begin{equation}
\setstretch{1.8}
\begin{array}{l}
m_h  = 123-127~{\rm GeV}
\\
m_{\tilde{g}} \geq 1.8~{\rm TeV} 
\\
M_{Z'} \geq 2.5 ~{\rm TeV} \\
0.8\times 10^{-9} \leq{\rm BR}(B_s \rightarrow \mu^+ \mu^-) 
  \leq 6.2 \times10^{-9} \;(2\sigma) 
\\ 
2.99 \times 10^{-4} \leq 
  {\rm BR}(B \rightarrow X_{s} \gamma) 
  \leq 3.87 \times 10^{-4} \; (2\sigma) 
\\
0.15 \leq \dfrac{
 {\rm BR}(B_u\rightarrow\tau \nu_{\tau})_{\rm MSSM}}
 {{\rm BR}(B_u\rightarrow \tau \nu_{\tau})_{\rm SM}}
        \leq 2.41 \; (3\sigma) \\
   0.0913 \leq \Omega_{{\rm CDM}}h^{2} \leq 0.1363~(5\sigma)     
\label{constraints}        
\end{array}
\end{equation}

We have emphasized the bounds on the Higgs boson \cite{:2012gk} and the gluino \cite{gluinoATLAS}, since they have drastically changed since the LEP era. One of the stringent bounds listed above comes from the rare B-meson decay into a muon pair, since the supersymmetric contribution to this process is proportional to $(\tan\beta)^{6}/m_{A}^{4}$. For solutions in the high $\tan\beta$ region in the fundamental parameter space $m_{A}$ needs to be large to suppress the supersymmetric contribution to ${\rm BR}(B_{s}\rightarrow \mu^{+}\mu^{-})$. Besides, the bound on the DM relic abundance is also highly effective to shape the parameter space, since the relic abundance of neutralino LSP is usually high over the fundamental parameter space. One needs to identify some coannihilation channels in order to have solutions compatible with the current WMAP and Planck results. The DM observables in our scan are calculated by micrOMEGAs \cite{Belanger:2006is} obtained by SARAH \cite{Staub:2008uz}. 

Among these experimental constraints, the most controversial one is that on the mass of $Z'$. The analyses within the UMSSM framework have set a bound on $M_{Z'}$ which can vary model dependently from about 2.7 TeV to 3.3 TeV \cite{Aaboud:2016cth}. Even though, these results were revealed recently, a new bound has just been released as $M_{Z'}\geq 4.1$ TeV \cite{ATLAS:2017wce}.  Such analyses are mostly based on the decay mode $Z' \rightarrow ll$, where $l$ can be either electron or muon with an assumption that $Z'$ decays mostly to these leptons; i.e. ${\rm BR}(Z'\rightarrow ll) \sim 1$. 

\begin{figure}[ht!]
\subfigure{\includegraphics[scale=0.4]{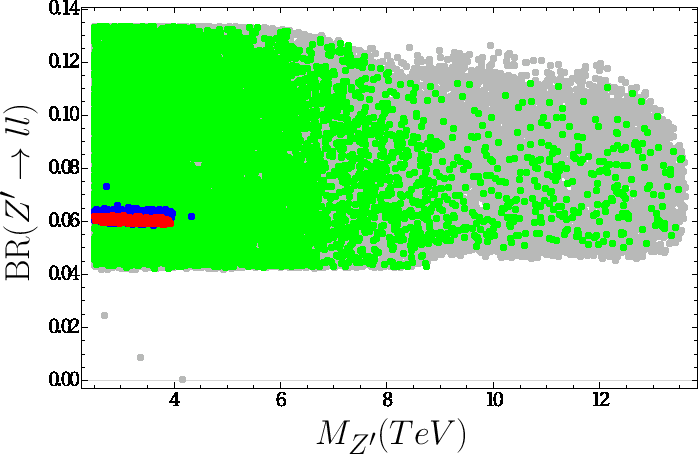}}
\subfigure{\includegraphics[scale=0.4]{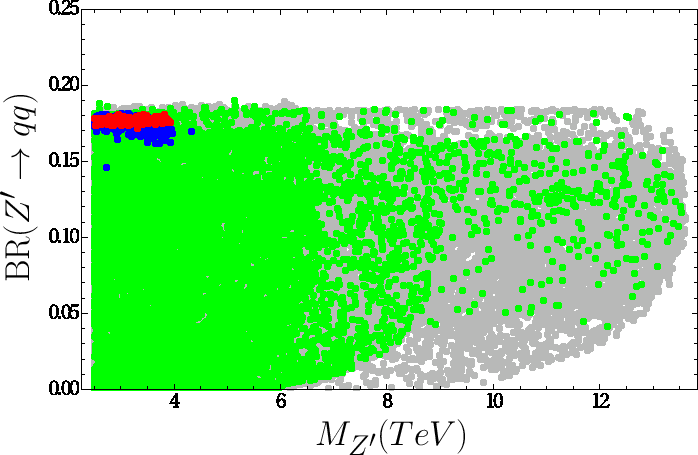}}
\caption{Decay modes of $Z'$ with largest branching ratio obtained in our analyses with plots in the ${\rm BR}(Z'\rightarrow ll)-M_{Z'}$ and ${\rm BR}(Z'\rightarrow qq)-M_{Z'}$ planes, where ${\rm BR}(Z'\rightarrow ll) \equiv {\rm BR}(Z'\rightarrow ee)+{\rm BR}(Z'\rightarrow \mu\mu)$, while $q$ in ${\rm BR}(Z'\rightarrow qq)$ denotes a final state quark from the first two families. All points are consistent with REWSB and neutralino being LSP conditions. Green points satisfy the LHC constraints listed above. Blue points form a subset of green, and they represent solutions for $\Delta m_{h} \leq 60$ GeV. Finally, red points are a subset of blue and they are consistent with the bound on the relic abundance of neutralino LSP within $5\sigma$ uncertainty. }
\label{fig:ZPdecays}
\end{figure} 

Figure \ref{fig:ZPdecays} shows the results obtained in our scans for the decay modes of $Z'$ with largest branching ratio obtained in our analyses with plots in the ${\rm BR}(Z'\rightarrow ll)-M_{Z'}$ and ${\rm BR}(Z'\rightarrow qq)-M_{Z'}$ planes, where ${\rm BR}(Z'\rightarrow ll) \equiv {\rm BR}(Z'\rightarrow ee)+{\rm BR}(Z'\rightarrow \mu\mu)$, while $q$ in ${\rm BR}(Z'\rightarrow qq)$ denotes a final state quark from the first two families. All points are consistent with REWSB and neutralino being LSP conditions. Green points satisfy the LHC constraints listed above. Blue points form a subset of green, and they represent solutions for $\Delta m_{h} \leq 60$ GeV. Finally, red points are a subset of blue and they are consistent with the bound on the relic abundance of neutralino LSP within $5\sigma$ uncertainty. The ${\rm BR}(Z'\rightarrow ll)-M_{Z'}$ plane shows that $M_{Z'}$ cannot exceed 4 TeV if one seeks less radiative corrections to the lightest CP-even Higgs boson (blue). This region also predicts ${\rm BR}(Z'\rightarrow ll) \sim 6\%$, which is far lower than the assumption behind the experimental analyses. In addition, considering the selected background processes in the analyses \cite{Aaboud:2016cth,ATLAS:2017wce}, the signal processes under consideration are those which involve with 4 leptons in their final states. In this case, the total branching ratio can be expressed in a good approximation as ${\rm BR}(Z'Z' \rightarrow 4l) \approx |{\rm BR}(Z'\rightarrow ll)|^{2}$, which provides more suppression for the results shown in the ${\rm BR}(Z'\rightarrow ll)-M_{Z'}$ plane.

According to our results, in the UMSSM framework constrained from the GUT scale, the largest branching ratio can be obtained for the decay modes yielding final states with hadrons. Our results show that ${\rm BR}(Z'\rightarrow qq) \sim 20\%$, as seen from the ${\rm BR}(Z'\rightarrow qq)-M_{Z'}$ plane. Even though it is large enough in comparison to those with leptonic final states, due to the uncertainties in the hadronic sector, such processes are not able to provide stringent bounds on $M_{Z'}$, yet. Even though, it is worth to be analyzed much deeper, it is beyond the scope of our work, and we set the lower bound as $M_{Z'}\geq 2.5$ TeV throughout our analyses. Such solutions can provide a testable phenomenology for $Z'$, and they can be excluded or confirmed by further analyses.

\section{Higgs Profile in UMSSM}
\label{sec:hprofile}

While the Higgs boson discovery is undoubtedly a breakthrough success for the SM, precise measurements are necessary to reveal  properties of the Higgs for which decay modes and couplings are also of crucial importance, since there is no direct signal for a new physics beyond the SM (BSM). Such measurements are also useful to distinguish the SM Higgs boson from those proposed by the BSM models. In the case of MSSM, although the heavier Higgs boson masses are at the decoupling limit ($m_{A}\gg M_{Z}$), and the lightest CP-even Higgs boson properties coincide with the SM Higgs bosons, MSSM can still yield some deviations in Higgs boson decay modes to the SM particles \cite{Bae:2015nva}. If such deviations are to be observed at the  experiments, then one can distinguish MSSM from the SM. In the UMSSM framework, the MSSM singlet field $S$, whose VEV is responsible for the $U(1)'$ symmetry breaking, can also mix with the MSSM Higgs doublets to form the lightest CP-even Higgs boson that is assumed to be SM-like. In this context, it might be important to distinguish such a Higgs boson from MSSM one using its properties. Such analyses can be performed with the effective Higgs couplings \cite{Bae:2015nva, Li:2016ucz} or equivalently through the branching ratios of the Higgs boson decay modes to the SM particles \cite{Carena:2001bg,Khachatryan:2016vau}. In our analyses we consider the branching ratios of the Higgs boson in comparison to the SM predictions in light of the current experimental measurements.

\begin{figure}[ht!]
\centering
\subfigure{\includegraphics[scale=0.4]{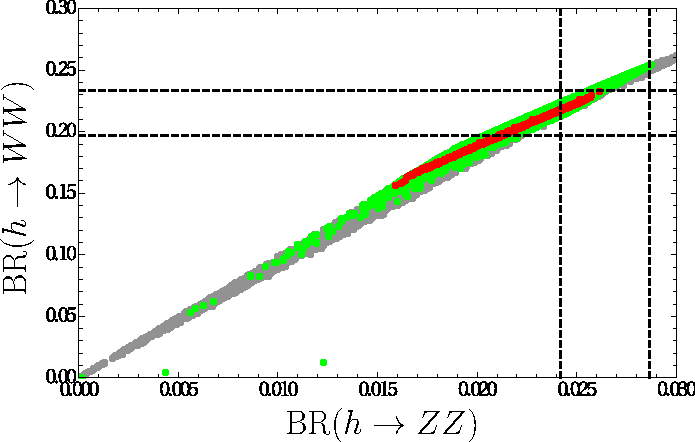}}
\subfigure{\includegraphics[scale=0.4]{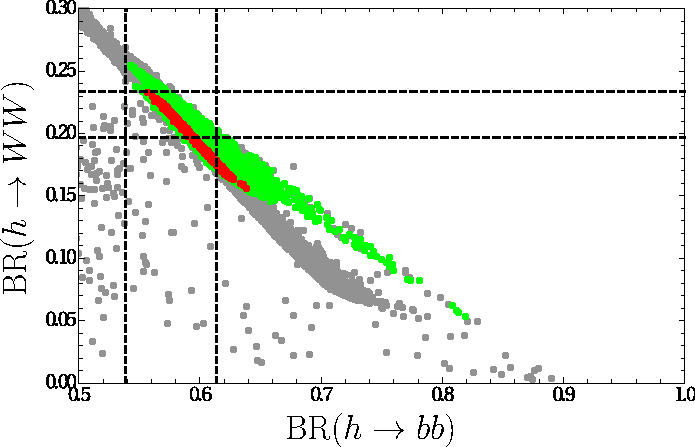}}
\caption{Higgs boson decays in the UMSSM framework with plots in the ${\rm BR}(h\rightarrow WW)-{\rm BR}(h\rightarrow ZZ)$ and ${\rm BR}(h\rightarrow WW)-{\rm BR}(h\rightarrow bb)$ planes. All points are consistent with the REWSB and neutralino being LSP. Green points represent the solutions allowed by the experimental constraints summarized in Sec. \ref{sec:scan}. Red points form a subset of green and they satisfy the DM bound on relic abundance of the LSP neutralino within $5\sigma$. The dashed lines indicate the SM predictions for the plotted decays within $1\sigma$ uncertainty.}
\label{fig:hWWZZbb}
\end{figure}

Figure \ref{fig:hWWZZbb} displays the Higgs boson decays in the UMSSM framework with plots in the ${\rm BR}(h\rightarrow WW)-{\rm BR}(h\rightarrow ZZ)$ and ${\rm BR}(h\rightarrow WW)-{\rm BR}(h\rightarrow bb)$ planes. All points are consistent with the REWSB and neutralino being LSP. Green points represent the solutions allowed by the experimental constraints summarized in Sec. \ref{sec:scan}. Red points form a subset of green and they satisfy the DM bound on relic abundance of the LSP neutralino within $5\sigma$. The dashed lines indicate the SM predictions for the plotted decays within $1\sigma$ uncertainty. Combined results from the ATLAS \cite{ATLAS:2014aga} and the CMS \cite{Chatrchyan:2013iaa} experiments yield ${\rm BR}(h\rightarrow WW) \approx 1.09 \times {\rm BR}(h\rightarrow WW)_{{\rm SM}}$ \cite{Khachatryan:2016vau}, where ${\rm BR}(h\rightarrow WW)_{{\rm SM}}$ stands for the SM prediction. Such an excess can be covered by the SM, if one considers its prediction for $h\rightarrow WW$ decay mode within about $2\sigma$ uncertainty band. However, as seen from the  ${\rm BR}(h\rightarrow WW)-{\rm BR}(h\rightarrow ZZ)$ plane of Figure \ref{fig:hWWZZbb}, the solutions allowed by the current experimental constraints including those from WMAP (red points) can only reach to the $1\sigma$ edge of the SM predictions for the $h\rightarrow WW$ decay. In this context, UMSSM predictions stay within the SM prediction region or below, but there is no solutions that can yield some excess in the $h\rightarrow WW$ decay mode. The deviation obtained from the ATLAS \cite{Aad:2014eva} and the CMS \cite{Chatrchyan:2013mxa} experiments is much larger for the $h\rightarrow ZZ$ decay mode that the combined results yield ${\rm BR}(h\rightarrow ZZ) \approx 1.29 \times {\rm BR}(h\rightarrow ZZ)_{{\rm SM}}$ \cite{Khachatryan:2016vau}. However, as in the case of $WW$ decay mode, UMSSM predictions for the $h\rightarrow ZZ$ barely stay within the close proximity of SM predictions. Many of the solutions predict  ${\rm BR}(h\rightarrow ZZ)$ smaller than the SM predictions and excluded if one insists to apply the SM predictions within $1\sigma$ uncertainty. 

Such lower predictions for the $WW$ and $ZZ$ decay modes can be explained with the mixing of the $S$ field with the MSSM Higgs doublets. Since this field is a gauge singlet, it does not interact with the $W-$ and $Z-$boson, and hence, its mixing in the SM-like Higgs boson lowers the predicted branching ratios in the $WW$ and $ZZ$ decay modes of the SM-like Higgs boson. Finally, we consider the $h\rightarrow bb$ decay in the ${\rm BR}(h\rightarrow WW)-{\rm BR}(h\rightarrow bb)$ plane as shown in Figure \ref{fig:hWWZZbb}. In contrast to the $WW$ and $ZZ$ decay modes the ATLAS \cite{Aad:2014xzb} and the CMS \cite{Chatrchyan:2013zna} experiments yield lower observation for the $h\rightarrow b\bar{b}$ decay mode as ${\rm BR}(h\rightarrow b\bar{b}) \approx 0.7 \times {\rm BR}(h\rightarrow b\bar{b})_{{\rm SM}}$ \cite{Khachatryan:2016vau}, which is way below the SM prediction. On the other hand, the UMSSM predicts ${\rm BR}(h\rightarrow b\bar{b}) \gtrsim 0.52$. 

The experimental measurements for some decay channels such as $h\rightarrow b\bar{b},\tau\bar{\tau}$ exhibit huge uncertainties and they can play a crucial role to constrain the new physics via the experiments conducted at the future colliders. While the uncertainty in these decay modes is stated with tens in percentage, it will be possible to reduce it to a few percent in the near future \cite{Atlas:2014susyCouplings}. Despite the uncertainties, the measurements in the $WW$ and $ZZ$ decay modes are well measured in comparison to other channels. These modes are also important, since some solutions may yield the lightest CP-even Higgs boson formed mostly by the MSSM gauge singlet $S$ field, which cannot be consistent with the assumption that the lightest CP-even Higgs boson is the SM-like Higgs boson in our analyses. In order to avoid such solutions, we will apply the SM predictions within $1\sigma$ as constraints on the CP-even Higgs boson decaying into the $W-$ and $Z-$ bosons.

Before concluding this section, we should also mention the loop induced decay mode of the Higgs boson into two photons. The experimental results for this decay mode indicate ${\rm BR}(h\rightarrow \gamma\gamma) \approx 1.14 \times {\rm BR}(h\rightarrow \gamma\gamma)_{{\rm SM}}$ \cite{Khachatryan:2016vau}. Although we do not present any plot for this decay, all the red points are consistent with the experimental constraints mentioned in Sec. \ref{sec:scan}, and they stay within the SM prediction region within $1\sigma$.

\section{Smaller Radiative Corrections}
\label{sec:R}

In this section, we consider the fundamental parameter space of UMSSM, which require low radiative corrections to the lightest CP-even Higgs boson consistent with the 125 GeV Higgs boson constraint. We quantify the values of these radiative contributions as $\Delta m_{h} \equiv \sqrt{m_{h_{{\rm loop}}}^{2}-m_{h_{{\rm tree}}}^{2}}$, which are defined in Eqs.(\ref{h0mass},\ref{h0loop}) in Sec. \ref{sec:hmass}. The least amount of the radiative corrections in the MSSM framework can be obtained as about 87 GeV \cite{Martin:1997ns}, and hence all solutions below this value can be advantageous of UMSSM. However, we consider only the solutions, which requires radiative corrections less than 60 GeV.

\begin{figure}[ht!]
\centering
\subfigure{\includegraphics[scale=0.4]{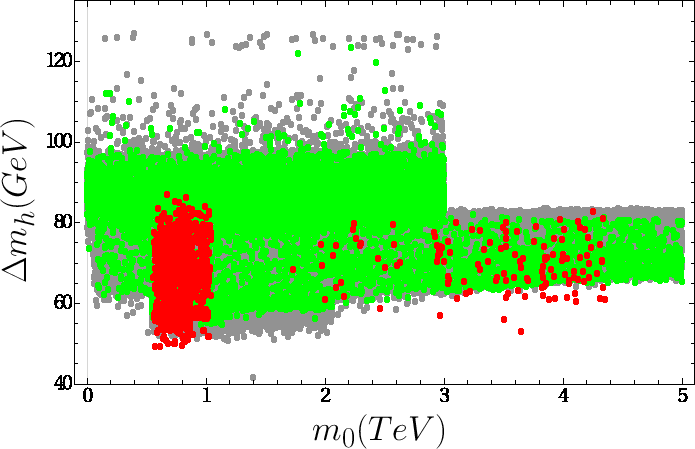}}
\subfigure{\includegraphics[scale=0.4]{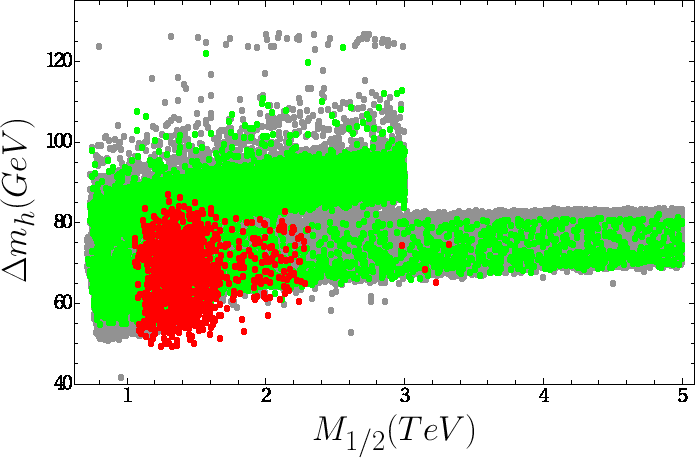}}
\subfigure{\includegraphics[scale=0.4]{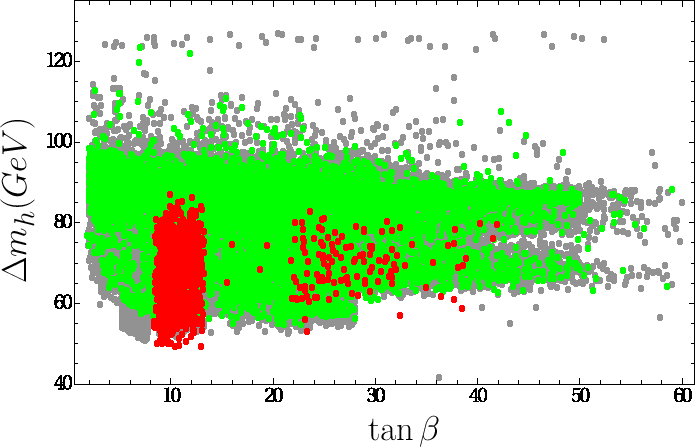}}
\subfigure{\includegraphics[scale=0.4]{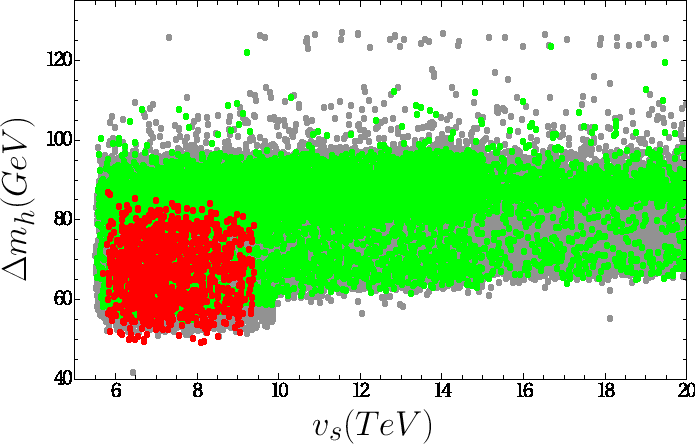}}
\caption{Plots in the $\Delta m_{h}-m_{0}$, $\Delta m_{h}-M_{1/2}$, $\Delta m_{h}-\tan\beta$ and $\Delta m_{h}-v_{S}$ planes. The colors have the same meaning as described for Figure \ref{fig:ZPdecays}, except the condition $\Delta m_{h} \leq 60$ GeV is not applied here. In addition, the green points also satisfy the SM predictions on ${\rm BR}(h\rightarrow WW)$ and ${\rm BR}(h\rightarrow ZZ)$.}
\label{loopfig1}
\end{figure}

Figure \ref{loopfig1} shows our results with plots in the $\Delta m_{h}-m_{0}$, $\Delta m_{h}-M_{1/2}$, $\Delta m_{h}-\tan\beta$ and $\Delta m_{h}-v_{S}$ planes. The colors have the same meaning as described for Figure \ref{fig:ZPdecays}, except the condition $\Delta m_{h} \leq 60$ GeV is not applied here. In addition, the green points also satisfy the SM predictions on ${\rm BR}(h\rightarrow WW)$ and ${\rm BR}(h\rightarrow ZZ)$. According to the results, it is possible to realize $\Delta m_{h}$ as low as about 50 GeV. Even though it mostly requires $m_{0} \lesssim 1$ TeV, as seen from the $\Delta m_{h}-m_{0}$ plane, it is possible to keep the radiative corrections low within whole range of $m_{0}$  TeV, although applying the dark matter constraint on the relic abundance of neutralino LSP restricts $m_{0} \lesssim 4$ TeV with good statistics. Similarly low values of $M_{1/2}$ tend to keep the radiative corrections low, even though the radiative corrections are still lower than those in the MSSM framework for $M_{1/2} \lesssim 3$ TeV consistently with the LHC constraints as well as the dark matter bound. These two parameters, $m_{0}$ and $M_{1/2}$, are important since they are effective in calculation of the stop and gluino masses at the low scale. Although the gluino mass is not directly effective in the lightest CP-even Higgs boson mass, it indirectly contributes, since it yields heavy stops through the loop effects. The $\Delta m_{h}-\tan\beta$ plane shows that there is no strong dependence on $\tan\beta$ in the radiative corrections, while the dark matter constraint allows only $\tan\beta \lesssim 45$. This is because there are also terms contributing to $\Delta m_{h}$ proportionally with $\cot\beta$ as seen in Eq.(\ref{h0loop}).  Finally we present the results for $v_{S}$, which determines the breaking scale of $U(1)'$ as well as $M_{Z'}$. The low radiative corrections require $v_{S} \lesssim 10$ TeV. This is also presenting the results in another way that $M_{Z'}$ cannot exceed 4 TeV in order to have the radiative corrections lower than 60 GeV as discussed in Sec. \ref{sec:scan}.

\begin{figure}[ht!]
\centering
\subfigure{\includegraphics[scale=0.4]{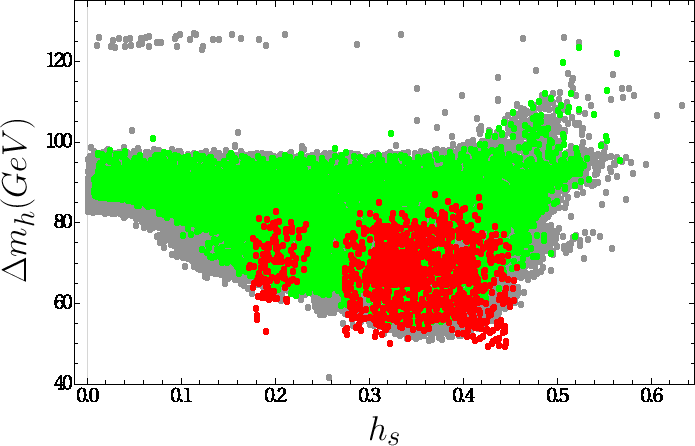}}
\subfigure{\includegraphics[scale=0.4]{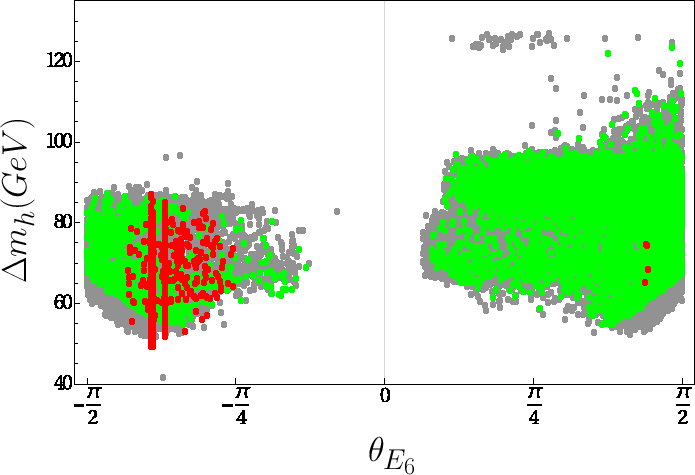}}
\subfigure{\includegraphics[scale=0.4]{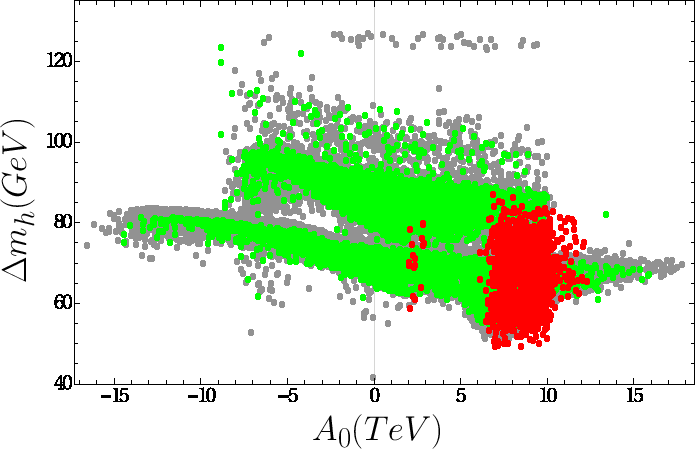}}
\subfigure{\includegraphics[scale=0.4]{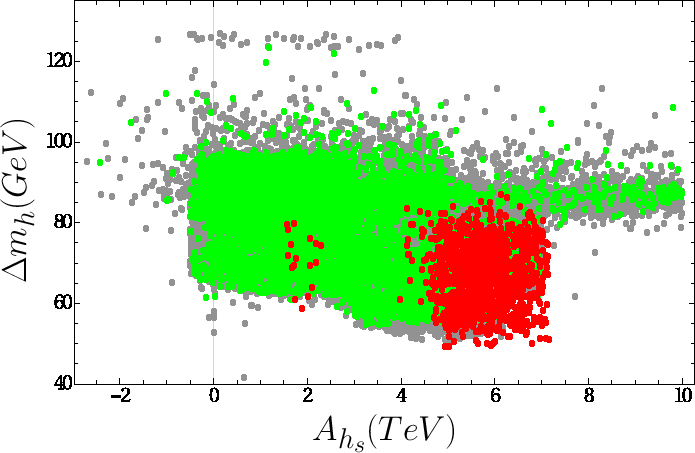}}
\caption{Plots in the $\Delta m_{h}-h_S$, $\Delta m_{h}-\theta_{E_{6}}$, $\Delta m_{h}-A_{0}$ and $\Delta m_{h}-A_{S}$ planes. The color coding is the same as Figure \ref{loopfig1}.}
\label{loopfig2}
\end{figure}

Figure \ref{loopfig2} displays our results for the other fundamental parameters of UMSSM with plots in the $\Delta m_{h}-h_S$, $\Delta m_{h}-\theta_{E_{6}}$, $\Delta m_{h}-A_{0}$ and $\Delta m_{h}-A_{S}$ planes. The color coding is the same as Figure \ref{loopfig1}. As seen from the $\Delta m_{h}-h_S$ plane, the radiative corrections tends to decrease with large $h_S$, and the lowest amount of radiative corrections can be realized for $h_S \lesssim 0.4$. As mentioned before, the radiative corrections exhibit also model dependency, which can be represented best with $\theta_{E_{6}}$, since this parameter yields different $U(1)'$ charge configurations. The $\Delta m_{h}-\theta_{E_{6}}$ shows that the lowest radiative corrections prefer the region with $1 \lesssim |\theta_{E_{6}}| \lesssim 1.5$, while the solutions consistent with the dark matter constraint mostly prefer the region with $\theta_{E_{6}} < 0$. The bottom panels of Figure \ref{loopfig2} represent the results in correlation with the trilinear scalar interactions terms $A_{0}$ (left) and $A_{S}$ (right). Since the accumulation of the solutions happens mostly in the low $\tan\beta$ region, as seen from Figure \ref{loopfig1}, these solutions require rather large $A$ terms, as $A_{0} \sim 7-10$ TeV and $A_{S} \sim 5-7$ TeV to satisfy the 125 GeV Higgs boson mass constraint. On the other hand it is possible to realize solutions with $A_{0}\sim 2$ TeV and $A_{S} \sim 2$ TeV, when $\tan\beta$ is large.

\begin{figure}[ht!]
\centering
\subfigure{\includegraphics[scale=0.4]{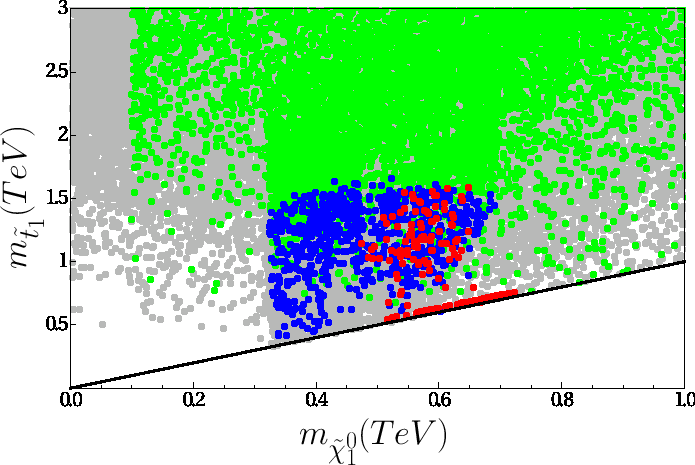}}
\subfigure{\includegraphics[scale=0.4]{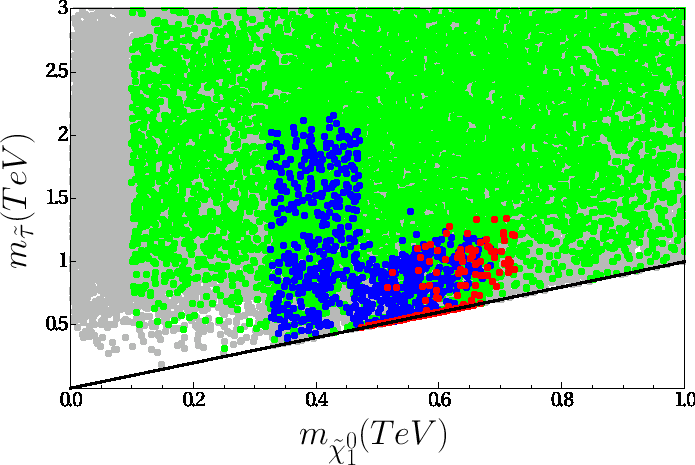}}
\subfigure{\includegraphics[scale=0.4]{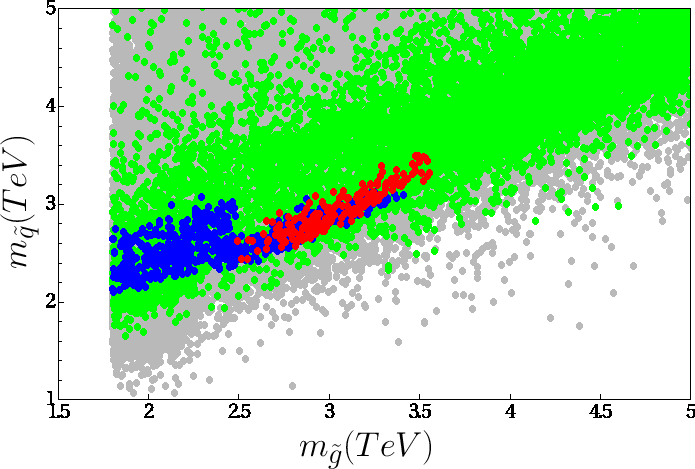}}
\subfigure{\includegraphics[scale=0.4]{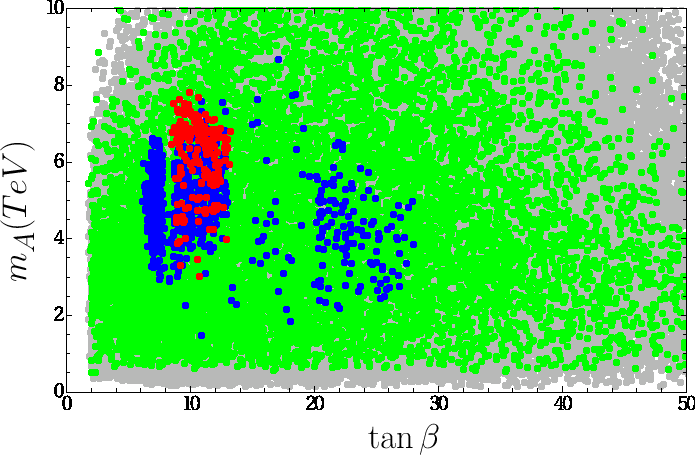}}
\caption{Plots in the $m_{\tilde{t}_{1}}-m_{\tilde{\chi}_{1}^{0}}$, $m_{\tilde{\tau}_{1}}-m_{\tilde{\chi}_{1}^{0}}$, $m_{\tilde{g}}-m_{\tilde{q}}$ and $m_{A}-\tan\beta$ planes. The color coding is the same as Figure \ref{loopfig1}. In addition the blue points represent solutions with $\Delta m_{h} \leq 60$ GeV, and red points form a subset of blue.}
\label{loopfig3}
\end{figure}

We consider the sparticle mass spectrum in Figure \ref{loopfig3} with plots in the $m_{\tilde{t}_{1}}-m_{\tilde{\chi}_{1}^{0}}$, $m_{\tilde{\tau}_{1}}-m_{\tilde{\chi}_{1}^{0}}$, $m_{\tilde{q}}-m_{\tilde{u}}$ and $m_{A}-\tan\beta$ planes. The color coding is the same as Figure \ref{loopfig1}. In addition the blue points represent solutions with $\Delta m_{h} \leq 60$ GeV, and red points form a subset of blue. The top panels reveal that the stop and stau can be degenerate with neutralino LSP when their masses are realized in $300-700$ GeV (blue). Applying the dark matter constraint on relic abundance of the neutralino LSP narrow this mass scale to $\sim 500-700$ GeV. Such solutions predict stau-neutralino and stau-neutralino coannihilation processes, which are responsible to reduce the relic abundance of neutralino LSP down to the ranges allowed by the dark matter constraint. The squarks of the first two families and gluino masses are always larger than about 1.5 TeV. The dark matter constraint restricts the masses of these sparticles further as $m_{\tilde{q}} \gtrsim 2$ TeV and $m_{\tilde{g}} \gtrsim 2.5$ TeV. Even though the mass bound on gluino is slightly larger ($m_{\tilde{g}} \gtrsim 1.9$ \cite{Sirunyan:2017cwe}) than what we applied in our analyses, the experimental constraints including those from dark matter automatically exclude the solutions which are not allowed by the current LHC results. The results for gluino with $m_{\tilde{g}} \gtrsim 2.5$ TeV provide also testable solutions in near future, since the next generation of colliders can probe the gluino mass up to about 3 TeV \cite{Baer:2016wkz}. The last plot in Figure \ref{loopfig3} represents the $A-$boson mass in the $m_{A}-\tan\beta$ plane. As it is seen, the results with low radiative corrections bound the mass scale of $A-$boson as $m_{A}\gtrsim 1$ TeV, and the dark matter constraint raises this bound up to about 4 TeV. These mass scales for $A-$boson are safely above the exclusion limit set as $m_{A}\gtrsim 1$ TeV \cite{Khachatryan:2014wca} for large $\tan\beta$.

\begin{figure}[ht!]
\centering
\subfigure{\includegraphics[scale=0.4]{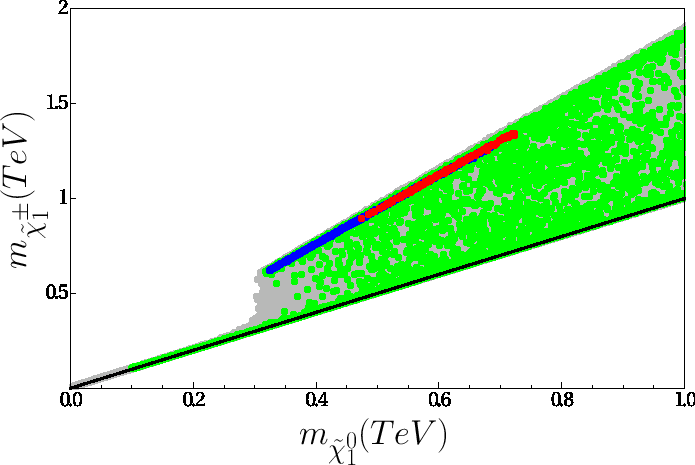}}
\subfigure{\includegraphics[scale=0.4]{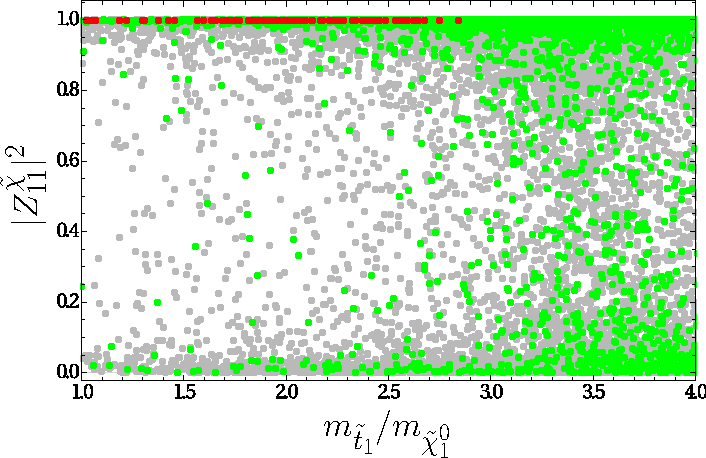}}
\caption{Plots in the $m_{\tilde{\chi}_{1}^{\pm}}-m_{\tilde{\chi}_{1}^{0}}$, ${\rm |Z^{\tilde{\chi}}_{11}|^2}-m_{\tilde{t}_{1}}/m_{\tilde{\chi}_{1}^{0}}$ planes, where $|Z^{\tilde{\chi}}_{11}|^2$ quantifies the percentage of the bino mixing in the dark matter formation. The color coding is the same as Figure \ref{loopfig3}.}
\label{loopfig4}
\end{figure}

Finally, we discuss the chargino and neutralino mass and comment about the dark matter formation in Figure \ref{loopfig4} with plots in the $m_{\tilde{\chi}_{1}^{\pm}}-m_{\tilde{\chi}_{1}^{0}}$, ${\rm |Z^{\tilde{\chi}}_{11}|^2}-m_{\tilde{t}_{1}}/m_{\tilde{\chi}_{1}^{0}}$ planes, where ${\rm |Z^{\tilde{\chi}}_{11}|^2}$ quantifies the percentage of the bino mixing in the dark matter formation, since the LSP neutralino is also assumed to be a candidate for the dark matter. The color coding is the same as Figure \ref{loopfig3}. The $m_{\tilde{\chi}_{1}^{\pm}}-m_{\tilde{\chi}_{1}^{0}}$ plane reveals the correlation between the LSP neutralino and the lightest chargino masses as $m_{\tilde{\chi}_{1}^{\pm}}\approx 2m_{\tilde{\chi}_{1}^{0}}$, when $\Delta m_{h} \leq 60$ GeV (blue). In this region the LSP neutralino mass is bounded at about 500 GeV from below by the dark matter constraint. Such a correlation between the chargino and neutralino masses also gives a hint about the dark matter formation. When the wino and/or higgsino are effective in dark matter formation, one usually obtains the relation $m_{\tilde{\chi}_{1}^{\pm}}\approx m_{\tilde{\chi}_{1}^{0}}$, since these supersymmetric particles also form the chargino mass eigenstates. However, the relation seen from the results indicates that these particles do not significantly mix in the dark matter formation; and hence the relic density of dark matter is saturated either by the bino or the singlino, the supersymmetric partner of the gauge singlet field $S$. The ${\rm |Z^{\tilde{\chi}}_{11}|^2}-m_{\tilde{t}_{1}}/m_{\tilde{\chi}_{1}^{0}}$ plane shows that the dark matter neutralino is pure bino, since its percentage in the dark matter formation is about $100\%$. These results can be concluded for the dark matter phenomenology as that the low $\Delta m_{h}$ regions in the fundamental parameter space of UMSSM yield pure bino dark matter. When a bino dark matter is scattered at nuclei, the cross-section of the process is usually low, since the dark matter interacts with nuclei through the hypercharge interactions. Thus, even the latest results of the LUX experiment \cite{Akerib:2016lao} do not provide strong impact on the direct detection predictions of the dark matter in this region.

\section{Notes on Fine-Tuning}
\label{sec:FT}

As discussed in Sec. \ref{sec:hmass}, the stop sector has a crucial role in realizing the consistent Higgs boson mass. In MSSM, a 125 GeV Higgs boson requires either stop masses at multi-TeV scale or large $A$ term \cite{Gogoladze:2009bd}. In the MSSM framework, large $A$ term worsens the required fine-tuning \cite{Demir:2014jqa}. On the other hand, one may expect the fine-tuning significantly improved, since the radiative corrections  to the Higgs boson mass do not have to be large. Even though the stop sector plays the main role in the consistent Higgs boson mass, they may not have to be very heavy, or have a large $A$ term. The minimization of the Higgs potential in UMSSM yields the following relation \cite{Athron:2013ipa} 

\begin{equation*}\hspace{-3.0cm}
\dfrac{M_{Z}^{2}}{2}=-\dfrac{h_{S}^{2}v_{S}^{2}}{2}+\dfrac{[(m_{H_{d}}^{2}+\Sigma_{d}^{d})-(m_{H_{u}}^{2}+\Sigma_{u}^{u})\tan^{2}\beta)}{\tan^{2}\beta -1}
\end{equation*}
\begin{equation}
+\dfrac{g_{Y}^{'2}(Q_{H_{d}}v_{d}^{2}+Q_{H_{u}}v_{u}^{2}+Q_{S}v_{S}^{2})}{2}\dfrac{(Q_{H_{d}}-Q_{H_{u}}\tan^{2}\beta)}{\tan^{2}\beta -1},
\label{FTUMSSM}
\end{equation}
%where $\mu \equiv h_{S}v_{S}/\sqrt{2}$ as defined earlier. $m_{H_{d}}$ and $m_{H_{u}}$ are the SSB mass terms for the MSSM Higgs doublets, while $\Sigma_{d}^{d}$ and $\Sigma_{u}^{u}$ are the radiative contributions to them respectively. $g_{Y}^{'2}$ is the gauge coupling associated with the $U(1)'$ gauge group, and $Q_{H_{d}}$, $Q_{H_{u}}$ and $Q_{S}$ are the charges of the fields indicated as subscripts under $U(1)'$. 

Even though Eq.(\ref{FTUMSSM}) does not exhibit an explicit dependence on the $A$ term, it contributes to the fine-tuning through the loops which are represented with $\Sigma_{d}^{d}$ and $\Sigma_{u}^{u}$, whose detailed calculations can be found in Ref. \cite{Baer:2012cf}. In MSSM, large radiative corrections result in worse fine-tuning. On the other hand, the second line of Eq.(\ref{FTUMSSM}) reveals the model dependency of the fine-tuning in the UMSSM frameworks, and it is possible to set a charge configuration for the fields such that they may reduce the effects of the large radiative corrections on the fine-tuning measurement. On the other hand, using Eq.(\ref{Zpmass}), the last term in Eq.(\ref{FTUMSSM}) can be expressed in terms of $M_{Z'}$. Substituting both $\mu$ and $M_{Z'}$ Eq.(\ref{FTUMSSM}) turns 

\begin{equation}
\dfrac{M_{Z}^{2}}{2}=-\mu^{2}+\dfrac{m_{H_{d}}^{2}-m_{H_{u}}^{2}\tan^{2}\beta}{\tan^{2}\beta -1}+\dfrac{M_{Z}^{2}}{2}\dfrac{(Q_{H_{d}}-Q_{H_{u}}\tan^{2}\beta)}{\tan^{2}\beta -1}.
\label{eq:MZ}
\end{equation}
where the loop contributions, $\Sigma_{d}^{d}$ and $\Sigma_{u}^{u}$, are now included in SSB masses $m_{H_{d}}^{2}$ and $m_{H_{u}}^{2}$ respectively. Following the usual definition in quantifying the fine-tuning \cite{Baer:2012mv} measure one can write 

\begin{figure}[ht!]
\centering
\subfigure{\includegraphics[scale=0.4]{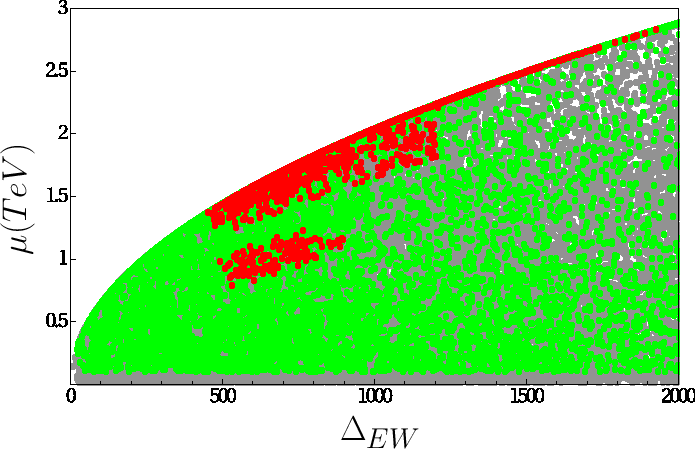}}
\subfigure{\includegraphics[scale=0.4]{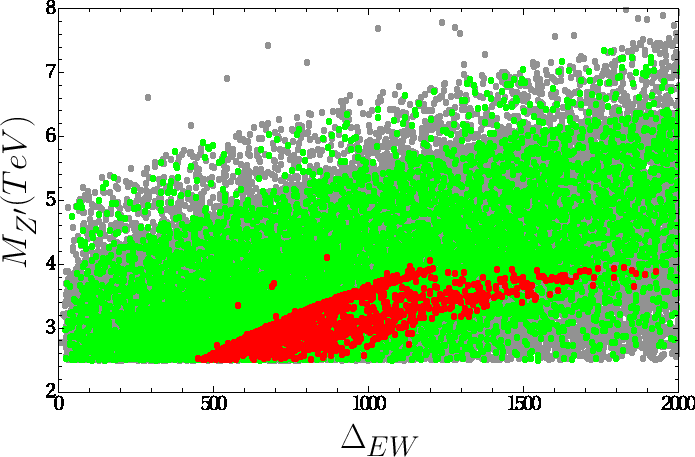}}
\subfigure{\includegraphics[scale=0.4]{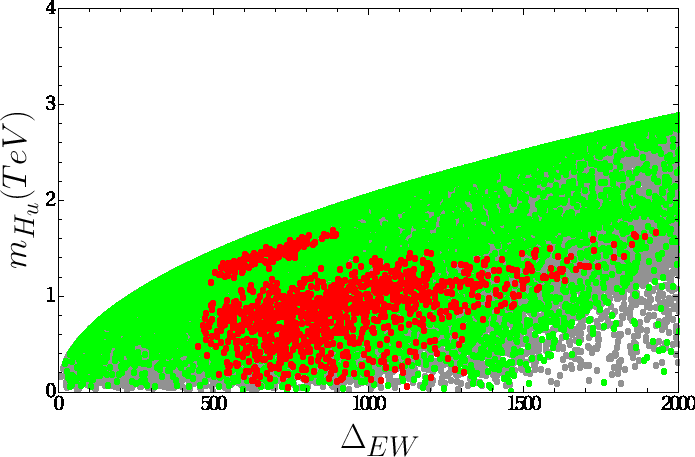}}
\subfigure{\includegraphics[scale=0.4]{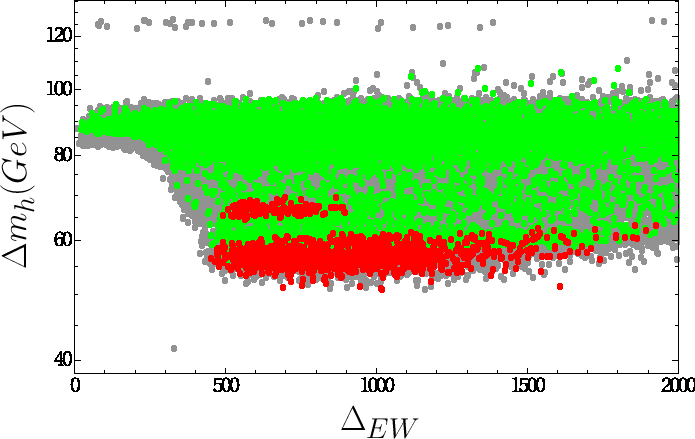}}
\caption{Plots for $\Delta_{EW}$ in correlation with $\mu$, $M_{Z'}$, $m_{H_{u}}$ and $\Delta m_{h}$. The color coding is the same as Figure \ref{fig:hWWZZbb}}
\label{fig:Delfine}
\end{figure}

\begin{equation}
\Delta_{EW}\equiv \dfrac{{\rm Max}(C_{i})}{M_{Z}^{2}/2}
\label{eq:delfine}
\end{equation}
with 

\begin{equation}
C_{i}=\left\lbrace 
\begin{array}{l}
C_{H_{d}}=\mid m^{2}_{H_{d}}/(\tan^{2}\beta -1) \mid \\ \\
C_{H_{u}}=\mid m^{2}_{H_{u}}\tan^{2}\beta/(\tan^{2}\beta -1) \mid \\ \\
C_{\mu}=\mid -\mu^{2}\mid \\ \\
C_{Z'}=\eval\dfrac{M_{Z'}^{2}}{2}\dfrac{(Q_{H_{d}}-Q_{H_{u}}\tan^{2}\beta)}{\tan^{2}\beta -1}\eval
\end{array}
\right. 
\label{eq:CFT}
\end{equation}
Here the impact of the heavy mass bound in $M_{Z}'$ can easily be seen. This impact can be suppressed in certain UMSSM models with $Q_{H_{d}},Q_{H_{u}}\sim 0$ selection. In such a case, the MSSM Higgs dublets become singlet under the $U(1)'$ gauge group, and the fine-tuning measure  more or less reduces to that obtained for MSSM \cite{DelleRose:2017ukx}. However, despite suppression in $C_{Z'}$, it does not remove the $M_{Z'}$ impact on the fine-tuning measure, since the heavy $M_{Z'}$ requires $v_{S}\gg v_{u,d}$. Namely, $v_{S}$ is also responsible for generating the $\mu-$term effectively, and its large values cause $\mu \gg \mathcal{O}(M_{Z})$ that leads to large fine-tuning again. Figure \ref{fig:Delfine} represents the results for $\Delta_{EW}$ in correlation with $\mu$, $M_{Z'}$, $m_{H_{u}}$ and $\Delta m_{h}$. The color coding is the same as Figure \ref{fig:hWWZZbb}. As seen from the $\Delta_{EW}-\mu$ plane, $\Delta_{EW}$ can be as low as 500, and in the general fashion of acceptable fine-tuning (say $\Delta_{EW}\leq 10^{3}$), such solutions can be considered in the acceptable fine-tuning region. However, $\Delta_{EW}$ raises quickly, and according to the results, mostly $\mu-$term is effective in measuring the fine-tuning. Similar behavior can be seen in the $\Delta_{EW}-M_{Z'}$ plane that the fine-tuning measure is becoming worse with heavy $M_{Z'}$ solutions. The results for $\mu$ and $M_{Z'}$ are reflection of the similar nature of $\mu$ and $M_{Z'}$ that is both of these parameters are induced effectively by $v_{S}$ for which one should  note that $v_{S}\gg v_{u,d}$.

The $\Delta_{EW}-m_{H_{u}}$ plane at the bottom of Figure \ref{fig:Delfine} shows that the MSSM relation $\mu \sim m_{H_{u}}$ does not have to hold; however, large $m_{H_{u}}$ values can still yield large fine-tuning predictions. Finally the $\Delta_{EW}-\Delta m_{h}$ displays radiative contributions to the Higgs boson mass and resultant fine-tuning. As seen from the results in this plane, the solutions with low radiative contributions may still yield large fine-tuning. Even though the fine-tuning measure can be interpreted in terms of the stop masses and $A$ terms, the low radiative corrections restrict such parameters to their relatively low values, and hence one might expect to have much lower fine-tuning measure. 

\begin{figure}[ht!]
\centering
\subfigure{\includegraphics[scale=0.4]{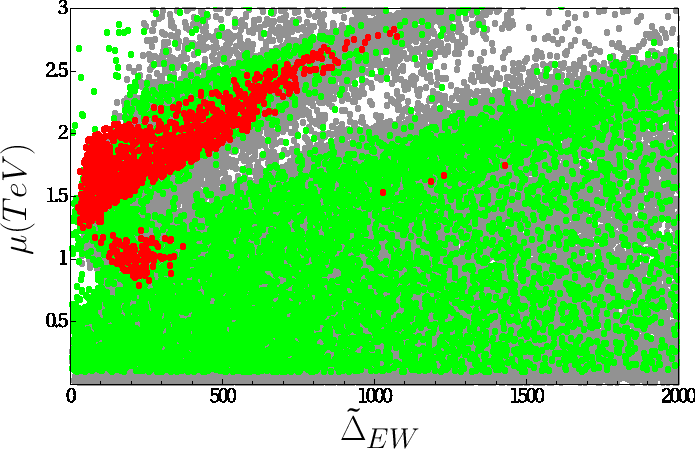}}
\subfigure{\includegraphics[scale=0.4]{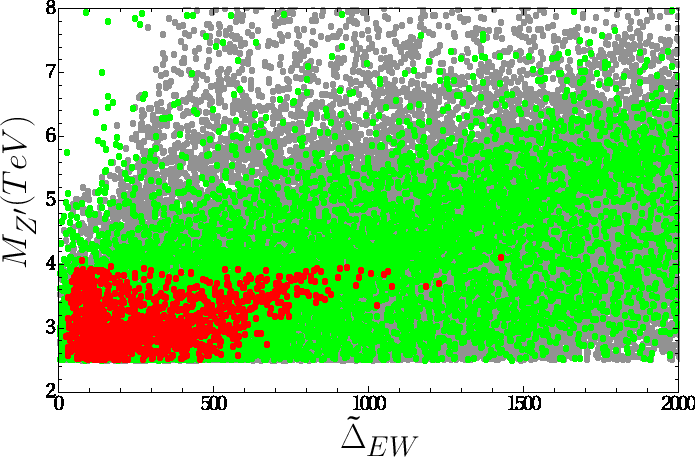}}
\subfigure{\includegraphics[scale=0.4]{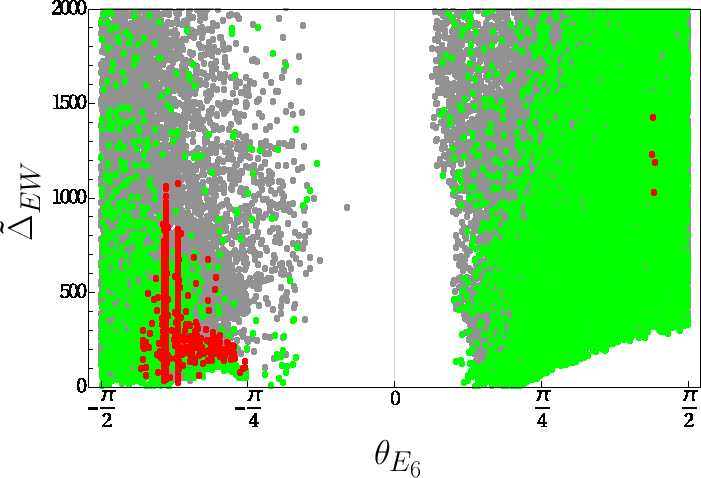}}
\subfigure{\includegraphics[scale=0.4]{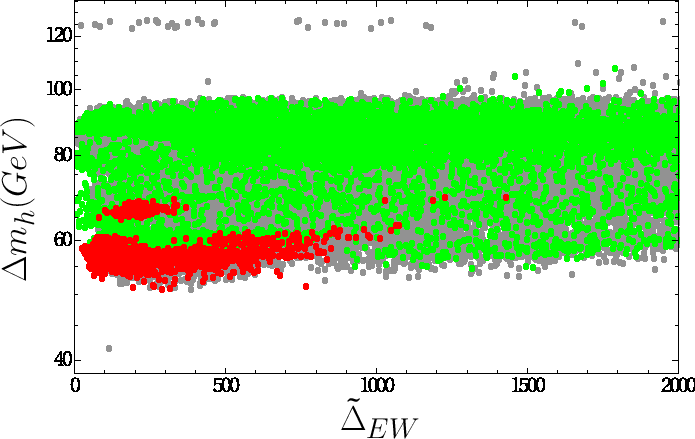}}
\caption{Plots for $\tilde{\Delta}_{EW}$ in correlation with $\mu$, $M_{Z'}$, $\theta_{E_{6}}$ and $\Delta m_{h}$. Only the solutions with $\Delta m_{h}\leq 80$ GeV are used in these plots. The color coding is the same as Figure \ref{fig:hWWZZbb}}
\label{fig:Delfinetilde}
\end{figure}   

These results might need to be reconsidered. since it is apparent from Eqs.(\ref{eq:delfine},\ref{eq:CFT}) that the $U(1)'$ breaking scale termed with $v_{S}$ is also the main factor that determines the fine-tuning measure at the electroweak symmetry breaking scale. On the other hand, the fundamental assumption behind the usual definition of $\Delta_{EW}$ is that the fine-tuning measure is determined by the cancellations among the parameters such as $\mu$, $m_{H_{u}}$ and $m_{H_{d}}$, which are, in principle, independent of each other, since they exhibit different nature. In this case, the large fine-tuning results shown in Figure \ref{fig:Delfine} may result from double counting, since $\mu$ and $M_{Z'}$ have more or less the same nature (when $v_{S}\gg v_{u,d}$) that both are induced by $v_{S}$ in the UMSSM framework. Let us rewrite Eq.(\ref{eq:MZ}) as

\begin{equation}
\dfrac{M_{Z}^{2}}{2}\approx-\tilde{\mu}^{2}+\dfrac{m_{H_{d}}^{2}-m_{H_{u}}^{2}\tan^{2}\beta}{\tan^{2}\beta -1}
\label{eq:MZtilde}
\end{equation}
with 
\begin{equation}
\begin{array}{l}
\tilde{\mu}^{2}= -\mu^{2}+\dfrac{M_{Z'}^{2}}{2}\dfrac{(Q_{H_{d}}-Q_{H_{u}}\tan^{2}\beta)}{\tan^{2}\beta -1} \\
\end{array}
\end{equation}
where we have neglected the terms with $v_{u}$ and $v_{d}$ in $M_{Z'}$ mass. If we define $\tilde{\Delta}_{EW}$ that is the fine-tuning measure in this approach, its definition will be in the same form as given in Eqs.(\ref{eq:delfine},\ref{eq:CFT}) except that $C_{\mu}$ needs to be replaced with $C_{\tilde{\mu}}$ as $\mu$ is replaced with $\tilde{\mu}$. 

Figure \ref{fig:Delfinetilde} show the results for $\tilde{\Delta}_{EW}$ in correlation with $\mu$, $M_{Z'}$, $\theta_{E_{6}}$ and $\Delta m_{h}$. The color coding is the same as Figure \ref{fig:hWWZZbb}. The $\tilde{\Delta}_{EW}-\mu$ plane shows that the fine-tuning measure represented with $\tilde{\Delta}_{EW}$ can be much lower despite the $\mu-$term being large. Indeed, it is possible to realize $\tilde{\Delta}_{EW}\sim 0$ even when $\mu \gtrsim 1.5$ TeV. In addition, in our approach, $\tilde{\Delta}_{EW}$ remains almost flat in $M_{Z'}$ mass as seen in the $\tilde{\Delta}_{EW}-M_{Z'}$ plane. Apart from the red points, which are consistent with all the experimental constraints mentioned in Sec. \ref{sec:scan}, $\tilde{\Delta}_{EW}$ can reach large values in green region, despite low radiative corrections. It arises from the model dependency in the expressions given so far. The configuration of the $U(1)'$ charges of the particles is not unique and infinite number of different configurations can be obtained by varying $\theta_{E_{6}}$ as given in Eq.(\ref{Chmixing}). For some values of $\theta_{E_{6}}$, especially $Q_{H_{u}}$ may lead $\tilde{\mu}> \mu$, while it yields $\tilde{\mu}<\mu$ for other $\theta_{E_{6}}$ values. The $\theta_{E_{6}}$ dependence is shown in the $\tilde{\Delta}_{EW}-\theta_{E_{6}}$ panel of Figure \ref{fig:Delfinetilde}. When $\theta_{E_{6}} \sim 0.5$, one can realize $\tilde{\Delta}_{EW}\sim 0$, and it raises when $\theta_{E_{6}} \gtrsim 0.5$. However, there is no red point with low fine-tuning measure in this region. Almost all red points with low fine-tuning are accumulated when $-1.4 \lesssim \theta_{E_{6}} \lesssim -0.8$. Finally, we also present the status of the fine-tuning with the radiative corrections to the Higgs boson mass to conclude that the low radiative correction solutions, in our approach, can be interpreted as those which form the low fine-tuning region in the fundamental parameter space of UMSSM. 

Before concluding this section  a few comments are useful. If we were to count  terms with $v_{d}$ and $v_{u}$ in $M_{Z'}$ as well as those with $v_{S}$, then they could be added to $m_{H_{d}}^{2}$ and $m_{H_{u}}^{2}$ in an appropriate way, but the results would be the same to  a good approximation, we checked this numerically. It should also be noted that the low fine-tuning measure in our approach, in contrast to the usual approach in MSSM, does not have to yield light Higgsinos at the low scale, which are quite interesting for the DM phenomenology.

\section{Conclusion}
\label{sec:conc}

 We consider the Higgs boson mass in a class of constrained UMSSM models and find that the amount of radiative contributions needed to realize a 125 GeV Higgs boson at the low scale can be as low as about 50 GeV, when $h_S$ is in the range $\sim 0.2-0.4$ and $v_{S}\lesssim 10$ TeV. Such low values of loop corrections needed to push the tree level predictions of the mass of the Higgs boson are not possible in MSSM whereas  as is NMSSM, UMSSM models need smaller loop induced corrections but in a model dependent way. Furthermore, because of the model dependency in predicting the Higgs boson mass, the  regions with relatively low radiative contributions prefer negative values of $\theta_{E_{6}}$ angle. In our study we observe  the least corrected UMSSM submodels reside near $\theta_{E_{6}}$ in [-$1.4, $-$0.8$]. 

In confronting the experiments, the lightest CP-even Higgs boson's decay modes are not obtained better than the SM predictions; thus, we restrict the solutions not to be worse than the SM in the Higgs boson properties. In this context, especially ${\rm BR}(h\rightarrow ZZ)$ provides the most stringent bound on the Higgs boson decays. In the mass spectrum of the supersymmetric particles, the region with low radiative contributions predict $m_{\tilde{t}} \lesssim 1.1$ TeV and $m_{\tilde{\tau}} \lesssim 2$ TeV. These sparticles can also be degenerate with the LSP neutralino in mass when they are lighter than about 700 GeV. The DM observations also restrict $m_{\tilde{t}}, m_{\tilde{\tau}} \gtrsim 500$ GeV. Such solutions also predict stop-neutralino and stau-neutralino coannihilation scenarions, which are effective in reducing the relic abundance of the LSP neutralino down to the ranges consistent with the current DM observations. The masses of the squarks of the first two families and gluinos lie from about 2 TeV to 3.5 TeV, and especially gluino solutions can be tested in the next generation of colliders. In addition, the CP-odd Higgs boson is found heavier than about 1 TeV, and its mass can be large up to 8 TeV in the regions consistent with the experimental constraints as well as being compatible with the requirement of low radiative contributions to the Higgs boson mass.  We find the lightest chargino can be as heavy as 1.2 TeV, but there is no solution which predicts degenerate chargino and neutralino LSP at the low scale. Hence, the DM is formed mostly by Bino, which yields low cross-section in scattering processes at nuclei. 
 
Finally we discuss the fine-tuning measure in the UMSSM framework, when the radiative contributions to the Higgs boson mass is low and all the experimental constraints are respected. In the usual definition, the fine-tuning measure is generally high and behaves worse over the fundamental parameter space. This situation can be explained by the heavy $M_{Z'}$ restriction. Such a heavy $Z'$ boson causes to high breaking scale for the $U(1)'$ symmetry, which is characterized with large $v_{S}$ values. In the usual definition, the required fine-tuning to realize the correct electroweak symmetry breaking scale is directly proportional to $v_{S}$; and hence, high $U(1)'$ symmetry breaking scales yield large fine-tuning predictions. Following this discussion, we reinterpreted the fine-tuning measure such that the effectively induced $\mu-$term and the contribution from $M_{Z'}$ can be combined into a single parameter, since they are induced by the same parameter; that is $v_{S}$. In such a redefinition, the fine-tuning measure can yield much lower values, even zero despite the heavy $M_{Z'}$ and large $\mu-$terms. The price for this redefinition is that the Higgsino DM solutions cannot be realized at the low scale, since the fine-tuning measure is not directly related to the $\mu-$term any more.

\vspace{0.3cm}
\noindent \textbf{Acknowledgement}
We would like to thank Shabbar Raza and \"{O}zer \"{O}zdal for useful discussions about the Higgs boson and $Z'$ properties. The work of \c{S}HT is supported by 2236 Co-Funded Brain Circulation Scheme (Co-Circulation) by The Scientific and Technological Research Council (TUBITAK) and the Marie Curie Action COFUND with grand no. 116C056.

\end{document}